\pdfoutput=1 

\documentclass{article}
\usepackage[utf8]{inputenc}
\usepackage{graphicx} 
\usepackage[numbers]{natbib}
\usepackage[T1]{fontenc}    
\usepackage{hyperref}       
\usepackage{amsmath,amsfonts,amssymb}
\usepackage{comment}
\usepackage[section]{placeins} 
\usepackage{multirow}
\usepackage{tabularx}
\usepackage{array}
\usepackage{pdfpages}
\usepackage{pdflscape}
\usepackage{longtable}

\title{Steering Generative Models for Protein Design: Aligning and Conditioning Strategies}
\date{-}
\author{Filippo Stocco$^{1,2,\diamondsuit}$, Michele Garibbo$^{1,\diamondsuit}$ and Noelia Ferruz$^{1,2,*}$ \\
\small$^{1}$Centre for Genomic Regulation, the Barcelona Institute of Science and Technology, \\
\small Dr Aiguader 88, Barcelona 08003, Spain\\
\small $^{2}$Universitat Pompeu Fabra (UPF), Barcelona, Spain\\
\small $^{\diamondsuit}$equally contributed\\
\small Emails: \texttt{\{filippo.stocco, michele.garibbo, noelia.ferruz\} @crg.eu}
}

\begin{document}
\maketitle
\begin{abstract}
Generative artificial intelligence models learn probability distributions from data and produce novel samples that capture the salient properties of their training sets. Proteins are particularly attractive for such approaches given their abundant data and the versatility of their representations, ranging from sequences to structures and functions. This versatility has motivated the rapid development of generative models for protein design, enabling the generation of functional proteins and enzymes with unprecedented success. However, because these models mirror their training distribution, they tend to sample from its most probable modes, while low-probability regions, often encoding valuable properties, remain underexplored. To address this challenge, recent work has proposed strategies for steering generative models toward user-specified properties. In this review, we survey and categorize these strategies, distinguishing approaches that modify model parameters, such as reinforcement learning or supervised fine-tuning, from those that keep the model's parameters fixed, including conditional generation, retrieval-augmented strategies, Bayesian guidance, and tailored sampling methods. Together, these developments are beginning to enable the steering of generative models toward proteins with desired properties.
\end{abstract}

\section{Introduction}
\begin{figure}[hbt]
     \centering
     \includegraphics[scale=0.20]{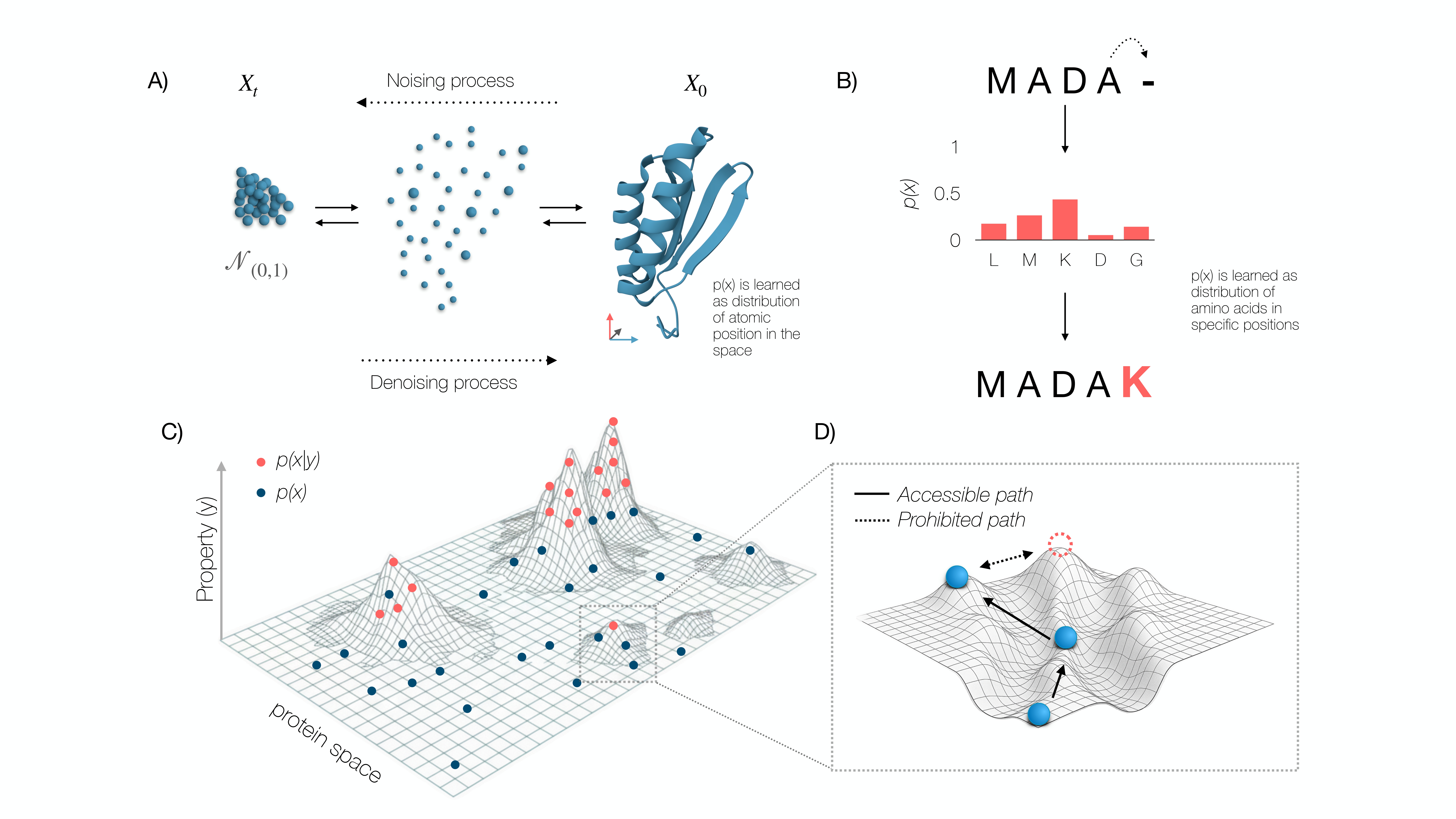}
     \caption{\textbf{Generative modeling approaches and schematic illustrations of protein fitness landscapes.} 
(A) \textit{Diffusion-based generative models.} Protein structures are represented by the distribution $p(x)$ of atomic coordinates in three-dimensional space. During training, true atomic coordinates $X_0$ are progressively corrupted by Gaussian noise until reaching a normal distribution $\mathcal{N}(0,1)$. At inference, the model reverses this process, iteratively denoising random noise to reconstruct a protein structure. 
(B) \textit{Sequence-based generative models.} Here, $p(x)$ denotes the distribution of amino acids across sequence positions. Training can proceed via masked-token prediction, in which the model infers the identity of masked residues, or via autoregressive next-token prediction, where sequences are generated one residue at a time. 
(C) \textit{Protein fitness landscape.} Schematic illustration of the relationship between the data distribution $p(x)$ (blue points), typically learned in an unsupervised manner, and the conditional distribution $p(y|x)$ (red points), which focuses sampling on regions associated with high fitness. 
(D) \textit{Evolutionary accessibility.} Evolution explores the fitness landscape through local, incremental mutations that can reach only contiguous high-fitness regions (solid paths), whereas transitions across fitness valleys (dashed paths) are inaccessible. In contrast, generative protein design models can, in principle, traverse the landscape in a less constrained manner, directly sampling from otherwise evolutionarily inaccessible regions.
}

     \label{fig:landscape}
 \end{figure}

Learning to generate samples from complex probability distributions lies at the core of modern generative modeling. In the context of proteins, the availability of large datasets has catalyzed the rapid development of powerful generative models (GM) for protein design. Some of the most adopted architectures are diffusion models, which reconstruct atomic coordinates by reversing a noise-adding process (Fig. \ref{fig:landscape}A)\cite{ahern_atom_2025, ingraham_illuminating_2023, stark_boltzgen_2025}, and protein language models (pLMs), trained either on masked sequence reconstruction (e.g. ESM \citep{lin2022language} or the MSA Transformer\citep{pmlr-v139-rao21a}) or on autoregressive next-token prediction (e.g. ProtGPT2\citep{ferruz_protgpt2_2022}, ProGen2-3\citep{bhatnagar2025scaling}, and Evo-1/2\citep{evo1, brixi2025genome}) (Fig. \ref{fig:landscape}B). In recent years, such models have enabled remarkable achievements, including the design of binders \citep{pacesa_one-shot_2025}, ligand-binding receptors \citep{rosettafoldallatom}, and \textit{de novo} enzymes \citep{braun_computational_2024} on unprecedented timescales. For broader overviews of generative AI for protein design, we refer the reader to recent reviews \cite{koh2025ai,yang2025illuminating,middendorf2026generative}.

Trained in a large corpus of natural sequences, structures, and functional annotations \citep[e.g.,][]{yang2025dayhoff,Cornman2024,bhatnagar2025scaling,varadi2024alphafold}, GMs can explore vast regions of the protein sequence–structure landscape \citep[e.g.,][]{bhatnagar2025scaling,chen2024xtrimopglm,cheng2024training,hayes2025simulating}.  However, protein engineering often targets exceptional properties that are rare or even disfavored by natural selection  (such as extreme thermostability in mesophilic organisms). These desirable functional optima may correspond to isolated high-fitness peaks separated by deep valleys in the evolutionary landscape (Fig. \ref{fig:landscape}C, D), representing trajectories that evolution is unlikely to traverse. Consequently, GMs trained solely on natural data tend to assign negligible probability to these regions, making them difficult to sample directly.  Further biases arise from uneven phylogenetic representation and residual annotation errors in the underlying datasets \citep[e.g.,][]{gordon2024protein,ding2024protein,Avasthi2024known,pugh_likelihood_2025}, which may further distort the learned landscape. In this sense, recent work highlight that simply increasing model scale does not guarantee monotonic improvements in fitness prediction \citep{hou2025understanding}. 

However, when appropriately guided, GMs have the potential to access low-probability yet functionally optimal regions of protein space. More formally, the protein design objective can be defined as generating protein sequences $x$ with desired properties $y$, where $y$ may represent attributes such as thermostability, catalytic efficiency, binding specificity, or other design goals. By contrast, the primary training objective of a GM is to model the distribution of natural proteins, $p(x)$.  To bridge this gap, recent approaches aim to model the conditional distribution $p(x|y)$, thereby biasing generation toward sequences with the desired attributes (Fig. \ref{fig:landscape}C). In many cases, this objective can also be viewed as an optimization problem, where the goal is to find $x$ that maximizes $p(x|y)$. Several recent techniques can be conceptually unified under this $p(x|y)$ framework, showing strong success in aligning GMs with specific design objectives. In this review, we categorize these strategies into two broad classes:
(i) parameter-updating alignment, which modify the model parameters so that the learned distribution $p(x)$ approximates $p(x|y)$, and
(ii) parameter-fixed steering, which guides generation toward target properties without altering the model’s underlying weights.

\subsection{Parameter-Updating Alignment Modifies weights to shift $p(x)$}
Parameter-updating alignment modify the underlying probability distribution by directly updating GMs' parameters. The most straightforward approach in this category is Supervised Fine-Tuning (SFT). In SFT, a pre-trained GM is further optimized under the same objective used during pre-training, but on a curated dataset of high-quality examples. For instance, a pre-trained protein language model (pLM) can be fine-tuned on a carefully assembled dataset from a target enzyme family to generate novel and distant members of that family. This process adapts the model’s parameters to a specific domain, effectively shifting its generative prior to align with the target data. SFT has achieved notable success, including the generation of enzymes \cite{munsamy_conditional_2024, madani2023large}, gene editors\cite{ruffolo_design_2024, ivancic_discovery_2025}, or bacteriophages \cite{king2025generative}.
\begin{landscape}
\begin{table}[h!]
    \centering
    \begin{tabular}{|>{\raggedright\arraybackslash}p{0.12\linewidth}|
                    >{\raggedright\arraybackslash}p{0.14\linewidth}|
                    >{\raggedright\arraybackslash}p{0.14\linewidth}|
                    >{\raggedright\arraybackslash}p{0.32\linewidth}|
                    >{\raggedright\arraybackslash}p{0.18\linewidth}|}
        \textbf{Category} & \textbf{Method} & \textbf{Intervention Local} & \textbf{Description} & \textbf{Examples}\\ \hline

        \multirow{2}{=}{Parameter-Updating Alignment} & Supervised finetuning (SFT) & \multirow{2}{*}{Model Weights} 
            & Fit the model to well-curated data, shifting the learned distribution towards the data.
            & \cite{munsamy_conditional_2024, madani2023large, ruffolo_design_2024, ivancic_discovery_2025}\\ \cline{2-2} \cline{4-5}

        & Reinforcement Learning & 
            & Aligns the model on feedback data over model’s outputs (preference or reward-driven learning).
            & \cite{widatalla_aligning_2024, stocco_guiding_2025, park2024improving, hayes2025simulating}\\ \hline

        \multirow{8}{=}{Parameter-Fixed Steering} & Prompt \& context programming & Prompting
            & Generation guided by structuring the input prompt with explicit instructions or templates.
            & \cite{alamdari_protein_2023, dauparas_robust_2022,dauparas_atomic_2025,cho2025protein}\\ \cline{2-5}

        & Retrieval-Augmented Generation (RAG) & Prompting
            & Enhances generation by dynamically incorporating external knowledge from a large corpus.
            & \cite{truong2023poet, weitzman_protriever_2025, truong2025understanding}\\ \cline{2-5}

        & Activation steering & Internal Representation
            & Direct manipulation of hidden states (e.g., residual stream) to promote or suppress attributes.
            & \cite{adams_mechanistic_2025, garcia_interpreting_2025, parsan_towards_2025}\\ \cline{2-5}

        & Output-dependent guidance & Decoding
            & Gradient-based guided generation in sequence space based on the inference output.
            & \cite{pacesa_one-shot_2025, cho_boltzdesign1_2025}\\ \cline{2-5}

        & Bayesian guidance & Decoding
            & Re-weights the probability distribution using Bayes’ theorem: $\tilde{p}(y\mid x)\propto p_\theta(y\mid x)\exp(\lambda s(y,x))$.
            & \cite{xiong2025guide}\\ \cline{2-5}

        & Sampling controls & Sampling
            & Alters sampling strategy (temperature, top-k, top-p, MCTS) to influence randomness.
            & \cite{brixi2025genome, darmawan2025sampling, ferruz2022protgpt2}\\ \hline
    \end{tabular}
    \caption{Overview of alignment and steering methods categorized by intervention stage within the generative process.}
    \label{tab:taxonomy_generative_updated}
\end{table}
\end{landscape}

While SFT effectively specializes GMs toward generating samples representative of a particular dataset, it does not provide the model with the ability to discriminate by data quality, i.e., to differentiate among varying degrees of a desired property. Similar limitations have been observed in Natural Language Processing (NLP), where SFT alone often leads to suboptimal alignment with user intent and \citep[][]{stocco_guiding_2025}, in some cases, catastrophic forgetting \citep[][]{shumailov2024ai}. Consequently, SFT is now commonly combined with reinforcement learning (RL) to achieve finer control over model behavior \citep[e.g][]{guo2025deepseek}.

In RL, a model learns to make optimal decisions by interacting with an environment through trial and error, receiving feedback in the form of rewards or penalties to maximize its cumulative reward over time. Unlike SFT, in RL the model is not provided with explicit examples of the desired outputs. Instead, the GM must infer and explore autonomously, potentially uncovering novel solutions that might not have been anticipated. More technically, a pre-trained model is treated as a policy $\pi_{\theta}(x)$ and updated to maximize a scalar reward, or the probability over preferences, while constraining excessive deviation from its pre-trained distribution. Similarly to SFT, this process enables a transition from the unconditional distribution $p(x)$ toward the desired conditional distribution $p(x|y)$. Today, RL is central to the alignment of large language models (LLMs) and has driven remarkable advances across diverse fields, from autonomous driving to game playing.

While different RL algorithms for model alignment have been proposed over the years, we can broadly group them in two main categories, 1) deep RL-based approaches and 2) direct preference learning. 
In deep RL-based approaches, a reward model is first trained on scored responses based on desired properties $y$ (e.g., human preferences). Subsequently, the reward model is used to update the policy (e.g., a pre-trained LLM), typically via a policy gradient update, shifting the underlying policy distribution $p(x)$ towards the properties $y$.
REINVENT (2017) \cite{olivecrona2017molecular} provides an early attempt to apply deep RL-based approaches to the molecular field.
REINVENT reframed the vanilla policy-gradient update (i.e., REINFORCE) as an "augmented likelihood objective". 
This objective enables adjusting the generative model parameters to favor samples with high scores on a target property $y$, while retaining probability mass near the original distribution $p(x)$, thereby preserving realism.
Shortly after, Proximal Policy Approximation (PPO, 2017) \citep[][]{schulman2017proximal} was introduced in the broader RL field, providing a key improvement over vanilla policy-gradient methods; its clipped surrogate objective approximates a trust region, guaranteeing more stable policy updates.
\begin{figure}[ht]
    \centering
    \includegraphics[scale=0.20]{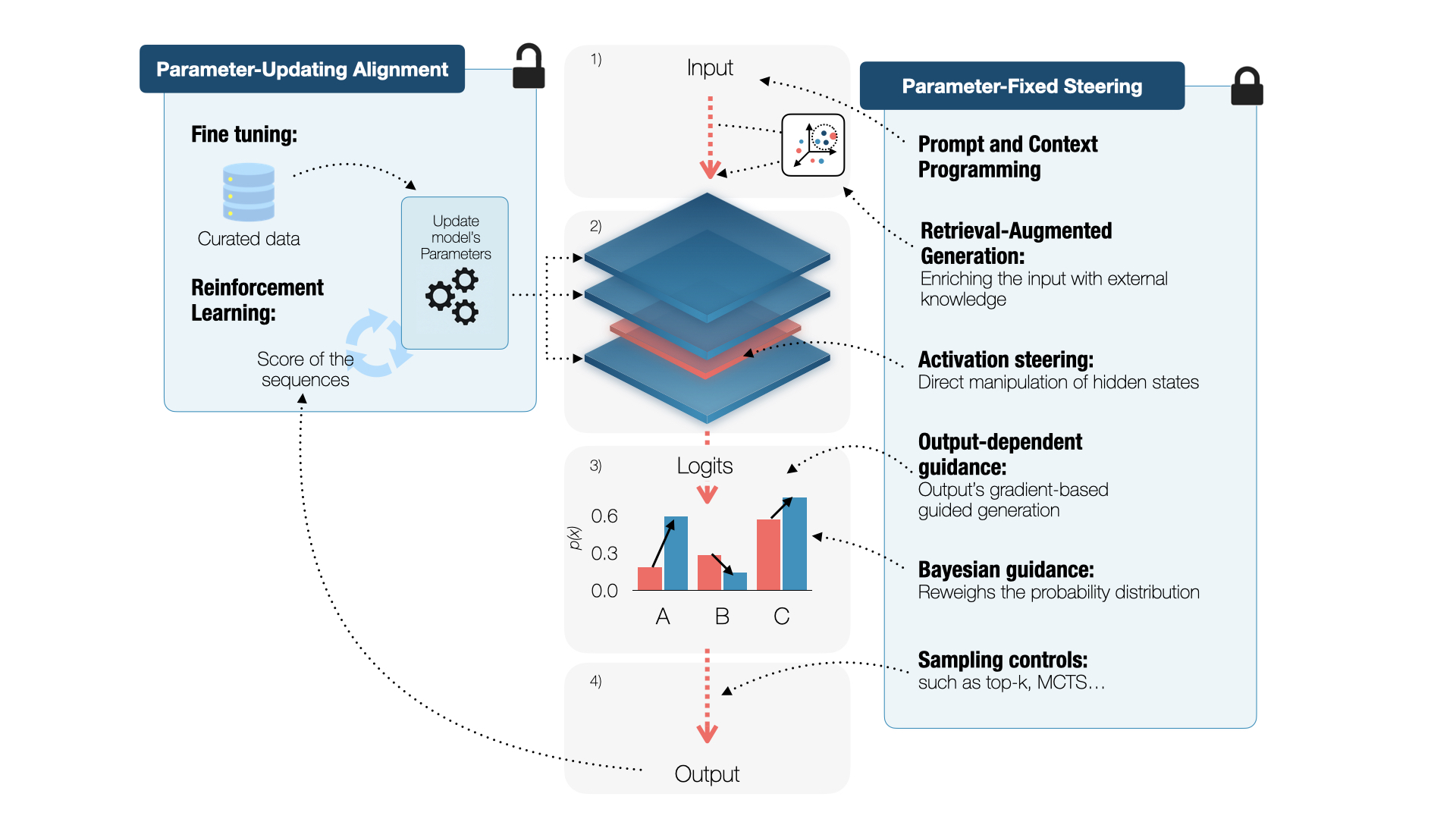}
    \caption{Different stages at which intervention can occur in protein generative modeling. Interventions may act on the model weights (Parameter-Updating Alignment), or without changing the base model parameters (Parameter-Fixed Steering). The latter category is further subdivided based on the stage of the model’s sampling pipeline where steering is applied: (1) specification of the input or context (e.g., prompting or retrieved context), (2) transformation into internal latent/activation representations, (3) decoding of latent representations into outputs, and (4) the sampling procedure that generates the final sequence or structure.}
    \label{fig:summary}
\end{figure}
The PPO update became the basis for RL from Human Feedback (RLHF) approaches \citep{ziegler2019fine}, allowing to align LLM to human preferences, while including a penalty for when the aligned model drifts too far from the original distribution $p(x)$. 
The more recent GRPO algorithm \citep[][]{shao2024deepseekmath} (2024) provides an alternative, where the reward is computed for a group of LLM's responses, augmenting the PPO update with a reward baseline (i.e., computed across the group), without needing to train any additional value model.
Direct Preference Optimization \cite{rafailov_direct_2024} (DPO, 2024) introduced the second category of RL approaches for model alignment: direct preference learning. DPO leverages the same preference signal as PPO-based methods, but cast alignment as a supervised objective over log-probability differences between ranked (human) preferences. In practice, this removes the need to train an explicit reward model, offering a simpler and potentially more stable route to preference alignment. These characteristics made DPO rapidly popular, and it has since spawned extensive variants of the original formulation.

These techniques have quickly met considerable success in the protein research realm, as summarized in table \ref{tab:rl_methods_protein}. Deep RL-based approaches have been used to update pLMs with experimental measurements; for example, RL from eXperimental Feedback (RLXF) aligns an ESM-based generator toward brighter CreiLOV variants while constraining drift from the original distribution\cite{blalock_functional_2025}. In another example,  Monte-Carlo Tree Search guided by a policy-value network (AlphaZero-like \cite{silver_mastering_2017}) has been applied to backbone construction, outperforming plain tree search on top-down design tasks\cite{renard_model-based_2024}. In parallel, direct preference learning has been used to bias structure-conditioned models toward stability (preferring stabilizing over destabilizing sequences given a target backbone), to improve output quality within specific protein families, and to reduce MHC-I epitope load while preserving the fold \cite{widatalla_aligning_2024, hayes2025simulating, gasser_tuning_2025}. Finally, recent frameworks aim to unify and extend these ideas: mutation-policy RL methods propose on-policy sequence edits under task-specific oracles \citep[][]{blalock_functional_2025}, while ProtRL generalizes the application of REINVENT, DPO and GRPO algorithms to protein engineering with pLMs, exemplified on the design of low-nanomolar EGFR inhibitors \cite{stocco_guiding_2025}.

While Direct preference learning approaches tend to be easier to work with compared to deep RL-based methods, it is still not clear which set of approaches lead to better performance under which conditions and, extensive benchmarking is needed \cite{yang_steering_2025}, especially, in relation to protein design applications.
Importantly, we believe the biggest limiting factor to the success of these methods in protein design is the difficulty of selecting good scoring metrics for most properties of interest.

\newcolumntype{L}[1]{>{\raggedright\arraybackslash}p{#1}}
\begin{table}[ht]
    \centering
    \begin{tabular}{L{0.25\linewidth}|L{0.50\linewidth}|L{0.25\linewidth}}
        \textbf{Method} & \textbf{Core objective} & \textbf{Protein-design uses (2022-2025)} \\ 
        \hline
        \textbf{REINFORCE}\newline\small{(Vanilla Policy Gradient, 1992 \citep[][]{williams1992simple})} 
        & Learns a policy such that actions with higher returns have higher likelihood of being sampled.
        &  AB-gen\citep[][]{xu2023ab} \\  
        \hline
        \textbf{PPO}\newline\small{(Proximal Policy Optimization, 2017\citep[][]{schulman_proximal_2017})} 
        & Plain policy gradients such as REINFORCE are quite noisy and unstable. PPO clips the model's gradient update to not deviate too much from original model distribution. 
        & RLXF\citep[][]{blalock_functional_2025}\citep[][]{wang2025proteinzero}\\  
        \hline
        \textbf{AlphaZero MCTS}\newline\small{(Monte Carlo Tree Search with policy–value networks, 2018 \citep[][]{silver2018general})} 
        & A neural networks with two heads is used to predict the most promising actions as well as the final rewards given a current position. This network is used to guide a MCTS to explore the most promising actions efficiently.
        & EvoPlay\citep[][]{wang_self-play_2023}, HighPlay\citep[][]{lin2025highplay} \\ 
        \hline
        \textbf{DPO}\newline\small{(Direct Preference Optimization, \citep[][]{rafailov_direct_2024})} 
        & Learns directly from ranked preference data,  without requiring an explicit reward model.
        & ProteinDPO\citep[][]{widatalla_aligning_2024}, ProtRL\citep[][]{stocco_guiding_2025} and Park et al.\citep[][]{park2024improving} \\  
        \hline
        \textbf{GRPO}\newline\small{(Group Relative Policy Optimization, \citep[][]{guo_deepseek-r1_2025})} 
        &For each prompt it samples a group of G candidate outputs, scores them with a reward, and uses relative (within-group) advantages to update the policy with a PPO-style clipped loss, without the need of a value model.
        & ProtRL\citep[][]{stocco_guiding_2025}, ProteinZero \citep[][]{wang2025proteinzero}\\  
        \hline
    \end{tabular}
    \caption{\textbf{Representative reinforcement learning frameworks applied to protein sequence design.} Summary of selected reinforcement learning (RL)–based approaches illustrating how core objectives have been adapted for protein or molecular design tasks between 2022 and 2025. The examples listed here are not exhaustive and focus on policy based RL, but highlight major methodological directions in the recent literature. For more details see \citep[][]{cao2025supervision}, and Supplementary.}
    \label{tab:rl_methods_protein}
\end{table}

\subsection{Parameter-fixed steering}
While parameter-updating alignment methods form a relatively well-defined class of methods, we use the term parameter-fixed steering to refer to a broader and more heterogeneous family of approaches. Despite their diversity, these methods share a common objective: conditioning model generation toward a desired property $y$ without modifying the generative model’s base parameters.
Conceptually, generation of learned GM can be decomposed into four stages: (1) an input or context specification stage (prompting), (2) the transformation of this input into internal latent or activation representations, and (3) the generation of the final output through the model’s decoding or (4) sampling procedure. The behavior of a GM can be influenced by intervening at any of these stages, as summarized in figure \ref{fig:summary}.

\paragraph*{Intervention at the input or context specification stage (prompting)}. In the first stage, input conditioning specifies desired attributes through context or control tokens. When such attributes are provided during training, the model effectively learns conditional generation (sampling from an implicit $p_\theta(x|y))$ rather than post-hoc steering of an unconditional $p_\theta(x)$. For example, several pLMs have been trained with various control tags, such as Enzyme Commission numbers (ZymCTRL,\cite{munsamy_conditional_2024}), Uniprot functional keywords (Progen1, \cite{madani2023large}), taxonomy (Evo1/2, \cite{evo1, brixi2025genome}), or combinations thereof \cite{yang_function-guided_2025}. \\
If the conditioning signal is not explicitly provided during training, desired constraints can instead be imposed at inference time via partial specification or hard constraints. For example, structure-conditioned sequence design models can fix selected residues (e.g., functional motifs) while redesigning the remaining positions \cite{dauparas_robust_2022}, and atomic-context conditioning can bias sequence design in the presence of ligands, nucleotides, or metals \cite{dauparas_atomic_2025}. More explicitly, some sequence generative models support infilling/inpainting, in which certain sequence segments are held fixed and the remaining residues are generated by masking and decoding the missing positions \cite{alamdari_protein_2023}. Others initialize sequences with unknown tokens (e.g., an all-X sequence) to induce structure hallucination in diffusion-based predictors, followed by iterative sequence redesign and structure re-prediction \cite{cho2025protein}.

Other approaches act directly on the latent encodings of protein sequences in a feedback-optimization loop, exploring regions of protein-structure space that satisfy user-defined design objectives such as ColabDesign, BindCraft and following methods \citep[][]{pacesa_one-shot_2025,cho_boltzdesign1_2025}. It is important to note that in machine learning "generative" strictly refers to models that sample from a learned probability distribution 
$p(x)$. By this definition, techniques based on activation maximization, where a predictor is coupled with gradient-based optimization to produce sequences (e.g., BindCraft), are not generative in the formal sense, as they do not define or sample from a prior. However, in the protein design community, these methods are often informally described as generative and, here,  we include such approaches  due to their prominent usage and success.

Alternatively, architectures such as BoltzGen\citep[][]{stark_boltzgen_2025} implement a conditional generative diffusion model with continuous guidance, in which encoded design conditions, such as binding-site specifications or structural constraints, are propagated throughout the denoising process to steer generation toward conformations consistent with the imposed design criteria \citep[][]{zhou2025cmadiff}.
With the aim to dynamically inject additional knowledge into the GM's context, Retrieval-augmented generation (RAG) has been successfully applied in differentcases\cite{truong2023poet,weitzman_protriever_2025,truong2025understanding}. By drawing upon semantically related examples from large databases, RAG enables more informed sampling and can enrich the design process with functional or structural priors not explicitly encoded in the pretrained model. A recent example is Protriever\cite{weitzman_protriever_2025}, which introduces a retrieval-augmented protein language model that jointly learns to retrieve homologous sequences and model their fitness, integrating evolutionary context at inference time without explicit structural supervision.
\paragraph*{Intervention at the hidden states.}
A different approach acts directly within the hidden states of the network, targeting the second stage. Activation steering manipulates internal representations, often within the residual stream, by injecting vectors that correspond to interpretable latent directions. Sparse autoencoders (SAEs) have been used to identify such interpretable features in protein language models, revealing latent dimensions correlated with properties like enzymatic activity, hydrophobicity, or thermostability \cite{adams_mechanistic_2025,garcia_interpreting_2025}. For example, Parsan et al. used SAE-derived features to bias structure predictions in ESMFold toward more hydrophobic conformations through feature steering\cite{parsan_towards_2025}, while Boxò et al. leveraged activity-associated features to steer ZymCTRL toward more active $\alpha$-amylases\cite{corominas2025sparse}.

\paragraph*{Intervention at the output stage and sampling.}

Finally, control can also be applied at the level of output. Bayesian guidance reweighs the probability distribution encoded by GMs using Bayesian principles, effectively combining the model’s prior with external evidence or predictive scores. Such strategies have been applied in protein design, where sequence likelihoods are updated according to functional predictors or activity \citep[][]{xiong2025guide}.
Similarly, sampling Controls manipulate the stochasticity of the GM's final output by, for example, manipulating "inference parameters" like temperature, top-k and top-p sampling, balancing diversity and fidelity, which is particularly important when sampling from vast sequence landscapes \cite{ferruz2022protgpt2}.
More advanced sampling techniques, such as beam search and Monte Carlo Tree Search (MCTS), allow to consider multiple GMs inference trajectories, selecting the optimal ones \citep[e.g.,][]{brixi2025genome, darmawan2025sampling}. 

Recent theoretical work underscores the generality of these approaches: flow matching in discrete state spaces has been shown to be equivalent to masked language modeling, autoregressive generation, and diffusion. 
This unifying perspective positions inference-time control as a suite of architecture-agnostic, plug-and-play techniques that can be ported across model classes with minimal modification \cite{xiong2025guide}.

\subsection{Conclusion and future prospects} 
Natural proteins are shaped by diverse and often competing forces: biochemical constraints, ecological pressure, and evolutionary mechanisms; making it difficult to define a single, global “fitness vector” that can be captured by GMs. 
Combined with well-known dataset biases \citep[e.g., see][]{gordon2024protein,ding2024protein,pugh_likelihood_2025}, these properties impose severe limitations on how much we can leverage the natural distribution of proteins to design and engineer proteins \textit{à la carte}, with full control over the design process.

Over the last two years, we have witnessed the emergence of several powerful methods for guiding GMs toward desired regions of the learned protein distribution (i.e., those encoding target properties for engineering). Here, we describe these methods and classify them into two broad categories: Parameter-Updating Alignment and Parameter-Fixed Steering.
These strategies have already been used to bias protein generative models toward brighter fluorescent proteins (e.g., CreiLOV variants)\cite{blalock_functional_2025}, increased stability\cite{widatalla_aligning_2024}, reduced MHC-I epitope load\cite{gasser_tuning_2025}, and improved target-binding designs, such as low-nanomolar EGFR inhibitors\cite{stocco_guiding_2025}.  

Despite this progress, several challenges still limit the immediate applicability of these approaches to fully autonomous protein engineering. A recurring issue across both categories is poor out-of-distribution (OOD) generalization. Parameter-Fixed Steering relies on the assumption that the pretrained model already encodes meaningful structure in the relevant regions of sequence space; consequently, steering cannot readily extrapolate beyond representations the base model has not learned. Parameter-Updating strategies offer greater flexibility by reshaping the learned distribution through fine-tuning, potentially extending coverage beyond pretraining. However, this flexibility comes at the cost of increased dependence on downstream data quality and optimization signals.

Many curated datasets used for supervised fine-tuning, as well as in silico scoring functions employed in reinforcement learning, remain rooted in the natural protein distribution and therefore inherit the same evolutionary biases and constraints that we ultimately attempt to overcome \citep[e.g.,][]{lin2023evolutionary,hayes2025simulating,dauparas_robust_2022}. In this context, lab-in-the-loop frameworks that iteratively combine generative models with experimental validation represent a particularly promising direction, as they could progressively enrich the data distribution \cite{vince_breaking_2025}. Complementary efforts aim to improve pretraining and optimization data by diversifying sequence sampling across the tree of life \cite{vince_breaking_2025} or introducing inductive biases that leverage evolutionary context through Multiple Sequence Alignments, enabling more efficient training \cite{akiyama_scaling_2025} and sampling \cite{pugh_likelihood_2025}.

A second major bottleneck concerns the availability of accurate and reliable scoring metrics. For many properties of interest—such as catalytic activity—robust predictive oracles remain difficult to obtain. Incorporating physics-based scoring methods, such as Rosetta \cite{alford2017rosetta} and FoldX \cite{schymkowitz2005foldx}, into reinforcement learning pipelines may help ground optimization in physically informed principles and reduce reliance on purely data-driven signals.

More fundamentally, it remains unclear whether alignment and steering methods genuinely uncover novel functional solutions or primarily recombine patterns already present in the training data. Addressing this question will require systematic and controlled empirical comparisons across diverse tasks and evaluation settings, allowing the field to rigorously assess the relative strengths and limitations of Parameter-Updating and Parameter-Fixed approaches\cite{yang_steering_2025, lu2025assessing}. As in many areas of biology, different strategies may prove optimal in different regimes—for example, de novo generation versus sequence optimization. Ultimately, advancing beyond the natural sequence distribution while maintaining reliability will be the defining challenge of the next phase of protein generative modeling, shaping the path toward truly controllable and new-to-nature protein design.

\subsection{Declaration of competing interest}
The authors declare no conflict of interest.

\subsection{Acknowledgments}
We thank Gerard Boxó and Alex Vicente and other members of the Ferruz lab for their thoughtful feedback. FS is supported by a fellowship from the “la Caixa” Foundation (ID 100010434, fellowship code DFI25-00620F). MG acknowledges funding from project CPP2022-009990 funded by MCIN/AEI/10.13039/501100011033 and by the European Union (FEDER, UE). NF acknowledges support from the Ramón y Cajal Fellowship (RYC2021-034367-I) funded by MICIU/AEI/10.13039/ 501100011033 and the European Union NextGeneration EU/PRTR, and from the European Union’s Horizon Europe programme under grant agreement No 101165231.
We acknowledge support of the Spanish Ministry of Science and Innovation through the Centro de Excelencia Severo Ochoa (CEX2020-001049-S, MCIN/AEI /10.13039/ 501100011033), and the Generalitat de Catalunya through the CERCA programme. We are grateful to the CRG Core Technologies Programme for their support and assistance in this work.  Funded by the European Union. Views and opinions expressed are however those of the author(s) only and do not necessarily reflect those of the European Union. Neither the European Union nor the granting authority can be held responsible for them. 

\bibliography{Bibliography}

\vspace{1em}
\noindent\normalfont\normalsize Papers of particular interest, published within the period of review, have been highlighted as: * of special interest ** of outstanding interest.
\newpage

\section{Supplementary}

\subsection{Reinforcement Learning Losses (Supplementary to Table~\ref{tab:rl_methods_protein})}
This section details the mathematical formulations of the reinforcement learning (RL) objectives summarized in Table~\ref{tab:rl_methods_protein}. Here $\pi_\theta$ denotes the policy, while the $D_{KL}(\pi_\theta||\pi_{ref})$ is a KL regularization term to penalize strong deviations from the reference model $\pi_{ref}$.

\paragraph{REINFORCE.}
The REINFORCE algorithm maximizes the expected reward by increasing the log-likelihood of sampled actions by their associated return \( r(x) \). The corresponding loss is:
\begin{align}
\mathcal{L}_{\mathrm{REINFORCE}} = - \mathbb{E}_{x \sim \pi_\theta} \left[r(x)\,\log \pi_\theta(x) \right]
\end{align}

\paragraph{Proximal Policy Optimization (PPO).}
PPO stabilizes training by preventing excessively large policy updates that may cause divergence. This is achieved by clipping the policy-ratio term $ r_i(\theta) = \pi_\theta(a_i|s_i) / \pi_{\mathrm{ref}}(a_i|s_i)$. The clipped objective is: 
\begin{align}
\mathcal{L}_{\mathrm{PPO}} = -\frac{1}{G}\sum_{i=1}^{G} \left[ \min \big( r_t(\theta)\,\hat{A}_t,\,\mathrm{clip}(r_t(\theta), 1-\epsilon, 1+\epsilon)\,\hat{A}_t \big) \right]\end{align} where $ \hat{A}_t $ denotes the advantage and $\epsilon$ defines the clipping range. While the advantage can be computed in different ways (e.g., GRPO-style), in PPO, it is classically computed relatived to a learned value function predicted the per-token expected reward. 
In RLHF, a KL penalty relative to a fixed reference model is added to the reward, ensuring that the new policy remains close to this reference.

\paragraph{AlphaZero.} Trains a single neural network with two heads, one that predicts the probability distribution over actions $p_\theta(a|s)$ and the other that predicts the expected sum of rewards given a state s $v_\theta(s)$. The training data comes from \textit{self-play} using a Monte Carlo Tree Search (MCTS) that produces the final real outcome $z$ and the used policies $\pi$ (i.e visit counts distribution for MCTS rollouts). The training loss is the following:

\begin{align}\mathcal{L}(\theta) = (z - v_{\theta}(s))^{2} \; - \sum_a\pi(a|s) \log p_\theta(a|s)\;+\; c\,\|\theta\|_{2}^{2}\end{align}

where $(z-v_{\theta}(s))^2$ is the value loss (mean square error between predicted and true value), $- \sum_a\pi(a|s) \log p_\theta(a|s)$ is the policy loss (cross entropy between MCTS policy and the network policy and $c\,\|\theta\|_{2}^{2}$ is a regularization term to avoid large weights updates (scaled by a constat $c$). 
\paragraph{Direct Preference Optimization (DPO).}
Introduced by Rafailov \textit{et al.}~(2023), DPO learns the optimal policy directly from pairwise preference data without requiring a reward model or value function. Given preferred $(x_w)$ and dispreferred $(x_l)$ samples, the objective is:
\begin{align}
\mathcal{L}_{\mathrm{DPO}} = -\mathbb{E}_{(x_w, x_l)} \!\left[\!
\log \sigma\!\big(\beta [\log \pi_\theta(x_w) - \log \pi_\theta(x_l)]\big)
\!\right],
\end{align}
where $\sigma$ is the sigmoid function and $\beta$ controls the strength of preference separation.

\paragraph{Group Relative Policy Optimization (GRPO).}
GRPO extends PPO by organizing responses into $|o|$ groups and computing advantages relative to each group’s mean performance, thereby reducing variance of each update and improving stability. 
It does not require a separate value model. 
The objective is:
\begin{align}
\mathcal{L}_{\mathrm{GRPO}} =
-\frac{1}{G}\sum_{i=1}^{G}\frac{1}{|o_i|}\sum_{t=1}^{|o_i|}
\min\!\Big(r_{i,t}(\theta)\,\hat{A}_{i,t},\,
\mathrm{clip}(r_{i,t}(\theta),1-\epsilon,1+\epsilon)\,\hat{A}_{i,t}\Big) + \beta\, D_{\mathrm{KL}}\!\left(\pi_\theta \| \pi_{\mathrm{ref}}\right)
\end{align}
where each group $o_i$ contains trajectories with similar characteristics. $\hat{A}_{i,t}$ denotes the group-relative advantage, which is computed as,
\begin{align}
\hat{A}_{i,t} = \frac{r_i - \text{mean}_t(r)}{\text{std}_t(r)}    
\end{align}
with the $\text{mean}_t$ and $\text{std}_t$ being computed over the group response.
\subsection{Retrieval-Augmented Generation}

Retrieval-Augmented Generation (RAG) augments a parametric policy $\pi_\theta$ with external knowledge retrieved from a database $\mathcal{D}$. Given an input $x$, a retriever $p_\eta(d|x)$ selects top-k relevant elements $d$ in the dataset $\mathcal{D}$ and marginalize the generator 
$\pi_\theta$.\begin{align} p_\theta(y|x) = \sum_{d \in \mathcal{D}}p_\theta(y|x, d)\, p_\eta(d|x) \end{align}
This formulation enables the model to combine parametric knowledge (stored in $\theta$) with non-parametric memory (stored in $\mathcal{D}$), improving factual grounding. In protein modeling settings, the database may consist of known sequences, structures, or functional annotations.

\subsection{Monte Carlo Tree Search}

Monte Carlo Tree Search (MCTS) is a planning algorithm that builds a search tree by iteratively simulating trajectories and updating value estimates.

At each state $s$ (e.g., a partially generated sequence), MCTS maintains a visit count $N(s)$, action visit counts $N(s,a)$, and action-value estimates $Q(s,a)$. During selection, actions are chosen using an upper confidence bound (UCB) criterion:
\begin{equation}
a^* = \arg\max_a \left[ Q(s,a) + c \sqrt{\frac{\log N(s)}{N(s,a)+1}} \right],
\end{equation}
where $c$ controls the exploration--exploitation trade-off.

After expanding a new node, a rollout (simulation) provides a reward $r$, which is backpropagated to update:
\begin{align}
N(s) &\leftarrow N(s) + 1, \\
N(s,a) &\leftarrow N(s,a) + 1, \\
Q(s,a) &\leftarrow \frac{N(s,a)\,Q(s,a) + r}{N(s,a)+1}.
\end{align}

When combined with policy--value networks (as in AlphaZero-style approaches), a learned prior $p_\theta(a|s)$ can guide exploration, improving search efficiency in large combinatorial spaces such as protein sequence design.

\subsection{Sparse Autoencoders}

Sparse Autoencoders (SAEs) are unsupervised models designed to learn compressed and interpretable latent representations. An encoder $f_\theta$ maps an input $x$ to a latent representation $z$, and a decoder $g_\phi$ reconstructs the input into $\hat{x}$. Encoder and decoder are trained with a reconstruction loss, which is usually the mean squared error:
\begin{align}\mathcal{L}_{\mathrm{rec}} = \|x - \hat{x}\|_2^2\end{align}
To encourage sparsity in the latent representation, a regularization term is added, typically an $\ell_1$ penalty:
\begin{align}\mathcal{L}_{\mathrm{SAE}} = \mathcal{L}_{rec} + \lambda\;\mathcal{L}_{aux}\end{align}
where $\lambda$ controls the sparsity strength.
SAE are commonly used for representation disentanglement, feature discovery, and interpretability analysis in large language and protein models. Ideally, SAE features are \textit{atomic}, meaning each latent dimension corresponds to a single, minimal, and semantically coherent factor (rather than an entangled mixture of multiple concepts), so features can be interpreted and recombined predictably.

\begin{landscape}
\begin{longtable}{|p{6.2cm}|p{8cm}|p{1cm}|p{2cm}|p{2cm}|}
\caption{Summary of works reported in this manuscript, with links to code repositories, when available.}
\label{tab:citations}\\
\hline
Title & GitHub & Reference & Type & GM \\
\hline
\endfirsthead

\hline
Title & Link & Reference & Type & GM \\
\hline
\endhead

\hline
\multicolumn{5}{r}{\small Continued on next page} \\
\hline
\endfoot

\hline
\endlastfoot

Conditional language models enable the efficient design of proficient enzymes
& -
& \cite{munsamy_conditional_2024}
& Fine-tuning
& pLM \\\hline

Large language models generate functional protein sequences across diverse families
& \href{https://github.com/salesforce/progen}{salesforce/progen}
& \cite{madani2023large}
& Fine-tuning
& pLM \\\hline

Design of highly functional genome editors by modeling the universe of CRISPR-Cas sequences
& \href{https://github.com/Profluent-AI/OpenCRISPR}{Profluent-AI/OpenCRISPR}
& \cite{ruffolo_design_2024}
& Fine-tuning
& pLM \\\hline

Discovery and protein language model-guided design of hyperactive transposases
& \href{https://github.com/Integra-tx/Piggybac_bioprospecting_pipeline}{Integra-tx/Piggybac\_bioprospecting\_pipeline}
& \cite{ivancic_discovery_2025}
& Fine-tuning
& pLM \\\hline

Aligning protein generative models with experimental fitness via Direct Preference Optimization
& \href{https://github.com/evo-design/protein-dpo}{evo-design/protein-dpo}
& \cite{widatalla_aligning_2024}
& Reinforcement Learning (DPO)
& pLM \\\hline

Guiding Generative Protein Language Models with Reinforcement Learning
& \href{https://github.com/AI4PDLab/ProtRL}{AI4PDLab/ProtRL}
& \cite{stocco_guiding_2025}
& Reinforcement Learning (DPO, GRPO)
& pLM \\\hline

Improving Protein Sequence Design through Designability Preference Optimization
& -
& \cite{park2024improving}
& Reinforcement Learning (DPO)
& Diffusion \\\hline

Simulating 500 million years of evolution with a language model
& \href{https://github.com/evolutionaryscale/esm}{evolutionaryscale/esm}
& \cite{hayes2025simulating}
& Fine-tuning
& pLM \\\hline

Protein generation with evolutionary diffusion: sequence is all you need
& \href{https://github.com/microsoft/evodiff}{microsoft/evodiff}
& \cite{alamdari_protein_2023}
& Prompt and Context Programming
& Diffusion \\\hline

Robust deep learning--based protein sequence design using ProteinMPNN
& \href{https://github.com/dauparas/ProteinMPNN}{dauparas/ProteinMPNN}
& \cite{dauparas_robust_2022}
& Prompt and Context Programming
& - \\\hline

Atomic context-conditioned protein sequence design using LigandMPNN
& \href{https://github.com/dauparas/LigandMPNN}{dauparas/LigandMPNN}
& \cite{dauparas_atomic_2025}
& Prompt and Context Programming
& - \\\hline

Protein Hunter: exploiting structure hallucination within diffusion for protein design
& \href{https://github.com/yehlincho/Protein-Hunter}{yehlincho/Protein-Hunter}
& \cite{cho2025protein}
& Prompt and Context Programming
& - \\\hline

PoET: A generative model of protein families as sequences-of-sequences
& \href{https://github.com/OpenProteinAI/PoET}{OpenProteinAI/PoET}
& \cite{truong2023poet}
& Retrieval-Augmented Generation
& pLM \\\hline

Protriever: End-to-End Differentiable Protein Homology Search for Fitness Prediction
& -
& \cite{weitzman_protriever_2025}
& Retrieval-Augmented Generation
& pLM \\\hline

Understanding protein function with a multimodal retrieval-augmented foundation model
& \href{https://github.com/OpenProteinAI/PoET-2}{OpenProteinAI/PoET-2}
& \cite{truong2025understanding}
& Retrieval-Augmented Generation
& pLM\\\hline

Mechanistic Interpretability of Antibody Language Models Using SAEs
& -
& \cite{adams_mechanistic_2025}
& Sparse Autoencoders
& pLM \\\hline

Interpreting and steering protein language models through sparse autoencoders
& \href{https://github.com/edithvillegas/plm-sae}{edithvillegas/plm-sae}
& \cite{garcia_interpreting_2025}
& Sparse Autoencoders
& pLM \\\hline

Towards Interpretable Protein Structure Prediction with Sparse Autoencoders
& -
& \cite{parsan_towards_2025}
& Sparse Autoencoders
& - \\\hline

One-shot design of functional protein binders with BindCraft
& \href{https://github.com/martinpacesa/BindCraft}{martinpacesa/BindCraft}
& \cite{pacesa_one-shot_2025}
& Output-dependent guidance
& - \\\hline

Boltzdesign1: Inverting All-Atom Structure Prediction Model for Generalized Biomolecular Binder Design
& \href{https://github.com/yehlincho/BoltzDesign1}{yehlincho/BoltzDesign1}
& \cite{cho_boltzdesign1_2025}
& Output-dependent guidance
& Diffusion \\\hline

ProteinGuide: On-the-fly property guidance for protein sequence generative models
& -
& \cite{xiong2025guide}
& Bayesian guidance
& Diffusion, pLM \\\hline

Genome modeling and design across all domains of life with Evo 2
& \href{https://github.com/arcinstitute/evo2}{arcinstitute/evo2}
& \cite{brixi2025genome}
& Fine-tuning
& gLM \\\hline

Sampling Protein Language Models for Functional Protein Design
& \href{https://github.com/i3LBI19-OATML/sampling_plm}{i3LBI19-OATML/sampling\_plm}
& \cite{darmawan2025sampling}
& Sampling controls
& pLM \\\hline

ProtGPT2 is a deep unsupervised language model for protein design
& \href{https://huggingface.co/nferruz/ProtGPT2}{https://huggingface.co/nferruz/ProtGPT2}
& \cite{ferruz2022protgpt2}
& Fine-tuning
& pLM \\\hline

Ab-Gen: Antibody Library Design with Generative Pre-trained Transformer and Deep Reinforcement Learning
& \href{https://zenodo.org/records/7657016}{https://zenodo.org/records/7657016}
& \cite{xu_ab-gen_2023}
& Reinforcement Learning
& pLM \\\hline

BoltzGen: Toward universal binder design
& \href{https://github.com/HannesStark/boltzgen}{HannesStark/boltzgen}
& \cite{stark_boltzgen_2025}
& Output-dependent guidance
& Diffusion \\\hline

CMaDiff: Cross-modal aligned diffusion for controllable protein generation
& \href{https://github.com/HPC-NEAU/PhysChemDiff}{HPC-NEAU/PhysChemDiff}
& \cite{zhou2025cmadiff}
& Reinforcement Learning
& Diffusion \\\hline

Self-play reinforcement learning guides protein engineering (EvoPlay)
& \href{https://github.com/melobio/EvoPlay}{melobio/EvoPlay}
& \cite{wang_self-play_2023}
& Reinforcement Learning
& Policy / Reinforcement Learning \\\hline

HighPlay: Cyclic peptide sequence design based on reinforcement learning and protein structure prediction
& -
& \cite{lin2025highplay}
& Reinforcement Learning
& Reinforcement Learning + structure predictor \\\hline

ProteinZero: Self-improving protein generation via online reinforcement learning
& -
& \cite{wang2025proteinzero}
& Reinforcement Learning
& pLM \\\hline

Functional alignment of protein language models via reinforcement learning (RLXF)
& \href{https://github.com/RomeroLab/RLXF}{RomeroLab/RLXF}
& \cite{blalock_functional_2025}
& Reinforcement Learning
& pLM \\\hline

\end{longtable}

\end{landscape}
\newpage

\end{document}